\documentclass{article}
\usepackage{ijcai24}

\usepackage{times}
\usepackage{soul}
\usepackage{url}
\usepackage[hidelinks]{hyperref}
\usepackage[utf8]{inputenc}
\usepackage[small]{caption}
\usepackage{graphicx}
\usepackage{amsthm}
\usepackage{booktabs}
\usepackage{algorithm}
\usepackage{algorithmic}
\usepackage{boldline}
\usepackage{multicol}
\usepackage{multirow}
\usepackage{subfigure}

\usepackage{amsmath,amsfonts}
\usepackage{array}
\usepackage{amssymb}
\usepackage{enumitem}
\usepackage{float}
\usepackage{makecell}
\usepackage{booktabs}
\usepackage{microtype}
\usepackage{textcomp}
\usepackage{stfloats}
\usepackage{verbatim}

\title{TNANet: A Temporal-Noise-Aware Neural Network for Suicidal Ideation Prediction with Noisy Physiological Data}

\author{
Niqi Liu$^1$\and
Fang Liu$^2$\and
Wenqi Ji$^1$\and
Xinxin Du$^1$\and
Xu Liu$^3$\and
Guozhen Zhao$^{3,4}$\thanks{Guozhen Zhao and Yong-Jin Liu are the corresponding authors.}\and
Wenting Mu$^5$
Yong-Jin Liu$^1$\footnotemark[1]
\affiliations
$^1$BNRist, Department of Computer Science and Technology, MOE-Key Laboratory of Pervasive Computing, Tsinghua University\\
$^2$State Key Laboratory of Media Convergence and Communication, Communication University of China\\
$^3$Multimodal Sensing and Computing Laboratory, Beijing, China\\
$^4$CAS Key Laboratory of Behavioral Science, Institute of Psychology\\
$^5$Department of Psychology, Tsinghua University
\emails
\{lnq22, jwq21\}@mails.tsinghua.edu.cn,
fangliu@cuc.edu.cn,
\{duxx, wmu, liuyongjin\}@tsinghua.edu.cn,
liuxu@zkpsych.com,
zhaogz@psych.ac.cn
}

\usepackage{color}
\newcommand{\lf}{\textcolor{black}}
\newcommand{\dxx}{\textcolor{black}}
\newcommand{\jwq}{\textcolor{black}}

\newcommand{\fixme}{\textcolor{black}}

\begin{document}

\maketitle

\begin{abstract}
The robust generalization of deep learning models in the presence of inherent noise remains a significant challenge, especially when labels are subjective and noise is indiscernible in natural settings. 
This problem is particularly pronounced in many practical applications\lf{. I}n this paper, we \lf{address} a special and important scenario of monitoring suicidal ideation, where time-series data, such as photoplethysmography (PPG), is susceptible to such noise.
Current methods predominantly focus on image and text data or address artificially introduced noise, neglecting the complexities of natural noise in time-series analysis. \lf{To tackle} this, we introduce a novel neural network model tailored for analyzing noisy physiological time-series data, \lf{named TNANet, }which merges advanced encoding techniques with confidence learning, enhancing prediction accuracy. 
% TODO-1: the dataset we built
Another contribution of our \lf{work} is the collection of a specialized \jwq{dataset of PPG signals} derived from real-world environments for suicidal ideation prediction. \fixme{Employing this dataset,} our TNANet achieves the prediction accuracy of \fixme{63.33}\% \fixme{in a binary classification task}, outperforming state-of-the-art models.
\lf{Furthermore}, comprehensive evaluations were conducted on three \lf{other} well-known public datasets with artificially introduced noise to rigorously test the TNANet's capabilities. These tests consistently demonstrated TNANet's superior performance by achieving an accuracy improvement of more than \fixme{10}\% compared to baseline methods.
\end{abstract}

\section{Introduction}
\label{sec:introduction}
In the realm of mental health research, particularly in predicting suicidal ideation, the presence of natural noise in physiological data presents a unique challenge due to the concealment~\cite{friedlander2012self}. The concealment of suicidal thoughts is prevalent across various demographics, such as in college students~\cite{burton2012college}, pilots~\cite{wu2016airplane}, and prisoners~\cite{liebling2012social}, which significantly hinders traditional assessment methods that rely on overt behavioral~\cite{laksana2017investigating} indicators or self-reported data \cite{nielsen2010pains}, complicating the detection and assessment processes.

\fixme{With the aforementioned noise, the task of analyzing physiological time-series data, such as photoplethysmography (PPG) signals, presents distinct analytical challenges.} PPG signals offer a more objective perspective on the dynamics of the autonomic nervous system, potentially bypassing the subjective barriers inherent in behavioral assessments~\cite{park2011autonomic}. However, the potential of PPG signals in detecting suicidal ideation has been largely underutilized, primarily due to the pervasive and complex issue of natural noise in data collection, \jwq{which is} a critical aspect that is often overlooked in computational research.

\textbf{\textit{Technical Challenge.}} Addressing the natural noise, especially prevalent in weakly annotated datasets from PPG signals, is pivotal in time-series data analysis. Unlike typical measurement errors or external interferences, this natural noise arises from the intrinsic complexities of physiological and emotional states, leading to ambiguous or misleading data points, thus posing significant challenges for accurate interpretation and analysis. \lf{Although recent studies~\cite{ma2023ctw,castellani2021estimating}}
have advanced the field of noisy time-series data analysis, they often neglect to fully address the distinct nature of noise found in physiological signals. Predominantly, these methodologies adapt strategies originally devised for image data, which may not entirely align with the specific nuances of noisy physiological time series. To bridge this gap, our study introduces a novel approach specifically tailored to the unique intricacies and complexities inherent in noisy physiological signals.

We present TNANet, a temporal-noise-aware neural network model crafted to adeptly handle natural noise in weakly annotated time-series data. Inspired by the success of EEGNet~\cite{lawhern2016eegnet} in processing physiological signals, TNANet synergistically combines self-supervised learning with a supervised convolutional framework. This integration allows TNANet to effectively extract and retain critical intra-class characteristics amidst noisy or ambiguous labels. By incorporating advanced encoding techniques and confidence learning, TNANet performs well in predicting suicidal ideation from PPG data, adeptly navigating the intricate landscape of natural noise.

In this study,  we focus on a specific population, prisoners. The phenomenon of suicidal ideation within prison populations presents a pressing challenge, \jwq{often serving} as a precursor to more acute suicidal behaviors and contributing significantly to overall suicidal risk~\cite{zhong2021risk}. In correctional settings, the prevalence of suicidal ideation is markedly higher than in the general population. This situation is exacerbated by the unique psychological and environmental stressors of prison life \cite{fazel2005suicides,opitz2013prison,fazel2017suicide}. These conditions create a complex landscape of mental health issues, further compounded by the prisoners' tendency to conceal their emotional distress.

\textbf{\textit{Contributions.}}
Our work makes significant strides in two aspects:
\begin{enumerate}
    \item We focus on the challenge of detecting suicidal ideation using peripheral physiological signals. Utilizing an affective reactivity paradigm, our study involved collecting PPG signals from real-world prisoners. We have delicately constructed a dataset for suicidal ideation prediction, encompassing both extracted features from these PPG signals and their corresponding labels for each individual.
    \item We propose a semi-supervised learning model that integrates a self-supervised encoding module with a supervised convolutional framework. This model excels in noisy data environments, employing a two-stage training strategy that first identifies noise through confidence learning, followed by a refinement phase with re-trained TNANet on the purified dataset.
\end{enumerate}

In the following sections, we delve into the specifics of TNANet's architecture, detail our experimental methodologies, present an in-depth analysis of our results, and discuss the far-reaching implications of our findings for the broader field of computational mental health.

\section{Related works}
\subsection{Suicidal Ideation Detection with Peripheral Physiological Signals}
Previous research has consistently demonstrated a correlation between blunted sympathetic nervous system (SNS) reactivity to psychological stressors and depressive states, often observed in individuals with major depressive disorder (MDD)~\cite{z14-salomon2013blunted,z11-brindle2017exploring,liu2023depressive}. Notably, these states are characterized by attenuated physiological responses, such as heart rate and blood pressure, during stress induction.

Recent studies have begun exploring the use of PPG signals concerning mental health, illustrating their potential in monitoring key physiological markers linked to mental states. For instance, PPG signals have been shown to effectively monitor heart rate variability~\cite{z8-allen2007photoplethysmography}, a crucial indicator of stress and emotional states, thus offering prospects for predicting psychological conditions or identifying emotional shifts~\cite{rinella2022emotion,lyzwinski2023use}. 

Transitioning from traditional methods to innovative approaches, the prediction of suicidal ideation has predominantly relied on subjective assessments, such as questionnaires and interviews~\cite{z4-liebling1995vulnerability,z1-favril2017suicidal}, or objective measures like electroencephalogram (EEG) and behavioral analysis~\cite{dolsen2017neurophysiological,laksana2017investigating}. These conventional techniques, while informative, often encounter limitations such as dependence on self-reporting accuracy or the necessity for complex equipment. PPG signals, in contrast, provide an affordable and ongoing monitoring alternative. Khandoker et al. have explored the association between PPG signals and suicidal ideation, particularly within the context of depressive symptoms~\cite{khandoker2017suicidal}. However, the direct application of PPG signals in predicting suicidal ideation using deep learning algorithms remains an unexplored area. Our study endeavors to fill this gap, advancing the understanding and potential of PPG signals in mental health monitoring.

\subsection{Computational Methods using \fixme{Biosignals}}
The domain of affective computing with physiological signals has seen a shift toward deep learning techniques,  
\fixme{which have been tailored for specific tasks in signal processing.}
In particular, EEG-based architectures have demonstrated adaptability for physiological signal analysis, leading to the exploration of similar methods for PPG signals~\cite{dominguez2020machine,zhu2020arvanet,siam2022portable}. Notably, neural architectures such as EEGNet~\cite{lawhern2016eegnet}, ShallowConvNet~\cite{schirrmeister2017deep}, and DGCNN~\cite{song2018eeg-dgcnn} have shown promising results in EEG analysis, and we are adapting them for PPG signal processing.

\subsection{Tackling Noise in Data Analysis}
\label{subsec:learn-noises} 
The issue of noisy labels in datasets, particularly in mental health applications, is a significant challenge. Recent approaches have focused on designing robust loss functions~\cite{ghosh2017robust,zhang2018generalized}, cultivating regularization techniques\lf{~\cite{zhu2020arvanet}}, and differentiating noisy data from clean data~\cite{bengio2009curriculum,kumar2010self-paced}. 
% co-teaching
Among them, Co-teaching~\cite{han2018co} employs dual classifiers, each learning from the other's most confident samples, enhancing noise resistance.
% Dividemix
DivideMix~\cite{li2020dividemix} treats noisy label learning as a semi-supervised problem, segregating data into clean and noisy sets for more effective model training.

In time-series data analysis, 
% SREA
SREA~\cite{castellani2021estimating} demonstrates robust handling of label noise in industrial contexts, offering insights applicable to similar challenges in physiological data analysis.
% CTW
In recent developments, CTW~\cite{ma2023ctw} represents a notable effort in addressing the noise through data augmentation tailored for time series data, primarily focusing on artificial noise.

% CL
Another type of approach is confidence learning~\cite{Northcutt2021CL}, which quantifies the likelihood of noise in samples by estimating the joint distribution between given labels and model outputs, serving as a preliminary filtering stage before formal training. 
% Encoder
Moreover, the use of self-supervised encoders, particularly Deep Belief Networks (DBNs)~\cite{smolensky1986information}, represents a significant advancement in this area. DBNs have shown their capability to reconstruct contaminated input data, offering a robust mechanism to counteract noise without reliance on accurate labels.

% Summarize
These methodologies collectively form a comprehensive foundation for our approach, where we incorporate these noise-handling techniques into TNANet for improved analysis of physiological time-series data.

\section{Development of a Specialized PPG Dataset}
\label{sec:data}
This study introduces a novel dataset, containing PPG signals collected from prisoners during an affective reactivity paradigm, as well as corresponding labels.

\subsection{Participants}
Male prisoners from a Hunan province prison in China participated voluntarily, excluding those in high security or hospitalized. The final group consisted of \fixme{2,190} right-handed prisoners, aged 40.96 $\pm$ 12.59.
 
\textbf{Ethical issues. }This study adhered to stringent ethical standards. Ethical clearance was obtained from the Institutional Review Board of the Institute of Psychology, Chinese Academy of Sciences. Participants voluntarily joined the study, fully informed about its nature, procedures, and their rights, including confidentiality and the option to withdraw at any time.

\subsection{Labels}
\label{sec:labels}
Labels were assigned to participants as `with suicidal ideation' (positive, coded as 1) or `without suicidal ideation' (negative, coded as 0), based on their suicide history, guard observations, and face-to-face psychological assessments. 
Guard ratings for each participant were collected using a one-question survey, where guards rated the likelihood of suicidal behavior on a 10-point Likert scale, with 1 indicating `not at all' and 10 being `very likely', based on the prisoner's daily behaviors.
The face-to-face assessments further evaluated factors contributing to an individual's suicidal ideation, including family support, loneliness, substance abuse, depression, and psychotic disorders.

\noindent\textbf{Positive Samples.}
\fixme{Positive labeling was contingent upon fulfilling three specific criteria concurrently: a documented history of suicide attempts or self-harm, a rating of 6 or above from prison guard assessments, and clinical confirmation through direct face-to-face evaluations.} This comprehensive approach identified 30 participants as positive cases.

\noindent\textbf{Negative Samples.}
For reliable negative labeling, participants with no suicide history, the lowest guard ratings \fixme{(1 on a 10-point Likert scale)} concurrently from at least three guards, and psychological verification were considered true negatives. This rigorous criterion segregated the negative samples into two subgroups: 21 true negatives and a larger uncertain group, where most participants were likely without suicidal ideation.

This purification procedure ensures a balanced number of true positive and true negative samples. Finally, the labels were divided into the following three groups: 30 true positives (\textit{TP}), 21 true negatives (\textit{TN}), and a larger group of uncertain negatives (\textit{UN}), ensuring a balanced representation of each category.
\lf{It is worth noting that most instances in \textit{UN} are true negatives.}

\subsection{\fixme{Stimuli and Apparatus}}
\label{para:stimuli}
A 5-minute video clip related to family affection was utilized as the evocative stimulus. The video depicted the reactions of a group of young adults upon witnessing their parents, who were artificially aged by 20 years through the use of special-effect makeup.
\fixme{The video clip is available online at \textit{\url{https://youtu.be/uu_NErqd9y8}}.}

\dxx{Prisoners watched video stimuli during the study, while we concurrently recorded their PPG signals using a custom wristband (Ergosensing, China). All prisoners wore the wristband on their non-dominant hand. The PPG signals were sampled at a frequency of 100 Hz.}

\subsection{Procedure}
Participants
%, grouped by cellblocks, 
were first acclimated with the wristbands used for PPG data collection. The \fixme{collection} involved two distinct phases: an initial 3-minute quiet sitting period for baseline data collection (`static' phase), followed by the `stimulation' phase where they watched the selected video clip. 
The amplitude of each timestamp in the `stimulation' phase was adjusted by subtracting the mean value of the `static' phase.

\subsection{Pre-processing}
\label{para: preProcessing}
In the preprocessing phase, personal identifiers were anonymized with unique codes.

\textbf{Feature Extraction and Normalization.} The PPG signals underwent a thorough process, starting with filtering using a 3rd-order band-pass Butterworth filter (0.6 Hz to 5 Hz range). These filtered signals were then segmented into 20-second windows, each overlapping 80\% with the next. Using the HeartPy~\cite{van2019analysing,van2019heartpy} Python package, different kinds of features like peak-to-peak interval and Shannon entropy were extracted, resulting in a dataset of 38 distinct features per window.
The full details of PPG features are listed in \fixme{Appendix 1}.

Signal durations, approximately 300 seconds, were divided into segments, each yielding 75 samples for PPG features. To standardize, sequences with accidental trigger errors were uniformly clipped to 70 windows, forming a $38 \times 70$ feature matrix for each sample. Min-max normalization was applied individually within each window to ensure consistent feature scales.

\textbf{Notations.} Each individual's feature sequence is represented as ${\mathcal{V}}\ [38, 70]$. Labels are designated with dual representation: $\tilde{y}_i \in \{0, 1\}$ for the assigned label of a sample, which may be inaccurate for some \textit{UN} samples, and $\vec{y^{}_i} \in \mathbb{R}^2$ indicating the output probability (where the first element indicates the probability of the sample being negative, and the second element indicates the probability of being positive) and  $y^{*}_i \in \{0, 1\}$ is used for the predicted label of sample $i$, aligned with the definitions in Section~\ref{sec:labels}, where:
\begin{equation}
y^{*}_i=
\left\{
    \begin{aligned}
        0, & \quad \vec{y^{*}_i}[0] \ge \vec{y^{*}_i}[1] \\
        1, & \quad otherwise 
    \end{aligned}
\right. 
\end{equation}

\section{Methodology}
\label{sec:dbnConvNet}
This study introduces the Temporal-Noise-Aware Neural Network (TNANet), designed to robustly predict suicidal ideation in noisy label environments. TNANet integrates Deep Belief Network (DBN) modules\lf{~\cite{Cai2016DBN}} with convolutional layers, enhanced by confidence learning for training with noisy data.

\begin{figure*}
    \centering
    \hspace{-5pt}
    \includegraphics[width=1.0\linewidth]{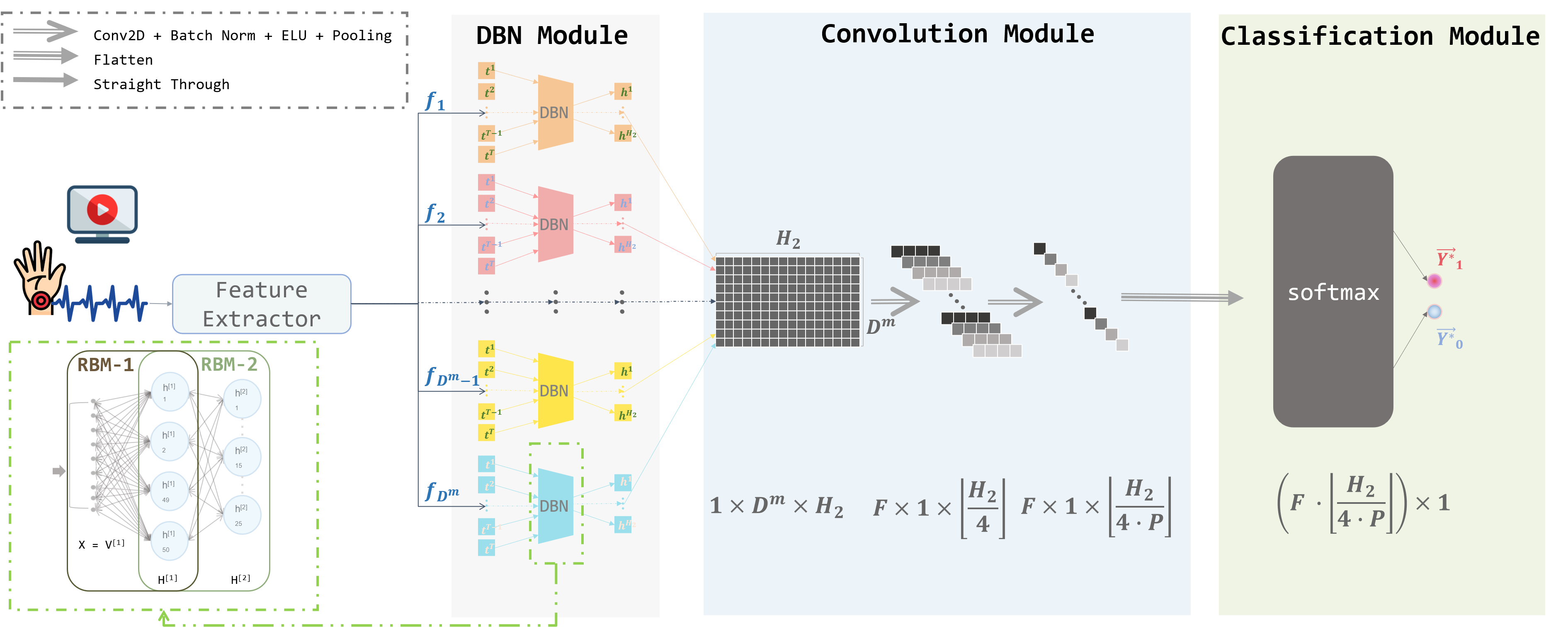}
    \caption{The architecture of our proposed TNANet model. It is composed of a DBN module, a convolution module, and a classification module. The input layer of the model is composed of $D^m$ vectors with a length $T$, where $D^m$ denotes the number of extracted features and $T$ represents the number of temporal points of each feature vector. The DBN module, which is constituted by RBM-1 and RBM-2, consists of one $T$-unit input layer ($V^{[1]}$), one 50-unit hidden layer ($H^{[1]}$), and one 25-unit output layer ($H^{[2]}$), with $H_2$ denoting the exact units amount. For each feature extracted from the preprocessed PPG data, the windows are formatted into a $T$-dim vector as the DBN input $X$. F is the number of filters (set to 16), P$=\min({H_2}//4, 8)$ is the pooling size in the second block of the convolution module. $\vec{Y^*}_1$ and $\vec{Y^*}_0$ are the predicted probabilities of being positive and negative, respectively.} 
    \label{fig: DBN_ConvNet}
\end{figure*}

\subsection{TNANet Architecture}
\label{Sec:Architecture}
As shown in Figure~\ref{fig: DBN_ConvNet}, TNANet consists of three parts: a DBN module, a convolution module, and a classification module.

The DBN module of TNANet comprises two cascading \lf{RBM components}. Each RBM is configured with $T=70$ input units (consistent with the number of windows), 50 units in the first hidden layer, and 25 units in the \lf{second hidden (}embedding) layer. This setup caters to the \fixme{38} extracted features \lf{described in Sec.~\ref{para: preProcessing}}, with each feature channel independently processed by a corresponding DBN instance.
The module first divides the input matrix into $D^{m} = 38$ sub-sequences, with each sequence forming a $T$-dimensional vector $\vec{v_i}$ for the $i^{th}$ feature. These vectors are then transformed into encoded vectors $\vec{h_i}$, as defined in Equation~\ref{eq: encode}, using the weight matrices ($\mathcal{W}{i1}$, $\mathcal{W}{i2}$) and biases ($b_{i1}$, $b_{i2}$) of the RBMs. This encoding process not only compresses the input data but also minimizes reconstruction loss (see Equation~\ref{eq: reconStructionLoss}), ensuring the retention of critical time series data components and effectively handling the variations among the features.
  
\begin{equation}
    \vec{h_i} = \mathcal{W}_{i2}(\mathcal{W}_{i1}\vec{v_i} + b_{i1}) + b_{i2}
    \label{eq: encode}
\end{equation}

The convolution module concatenates $\vec{h_i}$ back into a matrix $\mathcal{H}$ of size $[D^m, 25]$, aggregating the compact representations from different feature channels. Subsequently, a set of 2D filters (with dimensions $D^m \times 1$) are applied to these features to learn \jwq{the} feature weights and reduce the feature dimension.

The classification module, consisting of a fully connected layer and a softmax layer, outputs two probabilities, one for each label type. The final label for a sample is assigned based on the higher probability between these two outputs.
\begin{figure}[!ht]
    \centering
    \includegraphics[width=\linewidth]{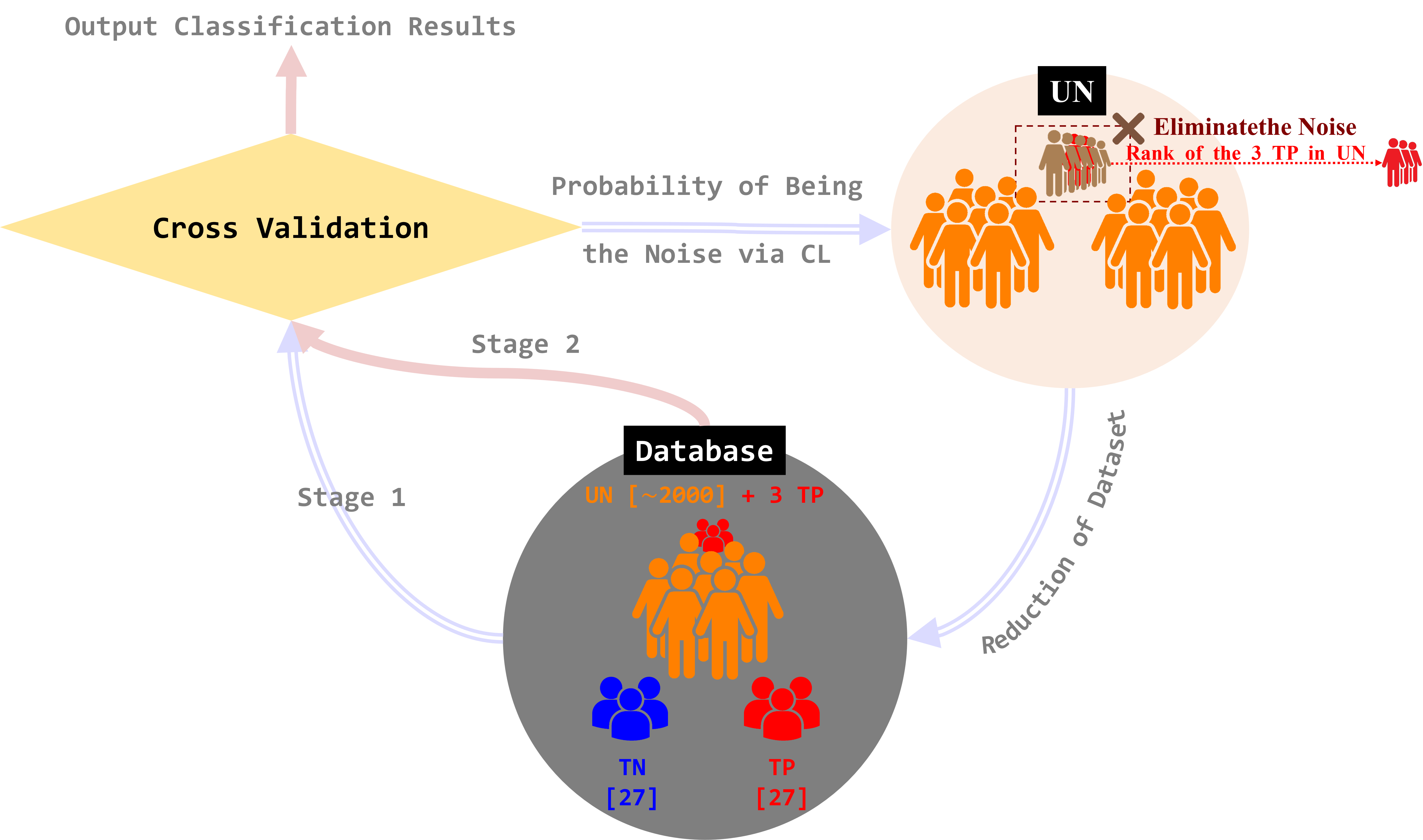}
    \caption{The training pipeline of TNANet. The purple arrows denote the first cross-validation stage and CL stage of TNANet, while the pink arrows denote the second cross-validation stage.}
    \label{fig:train_process}
\end{figure}

The detailed network structure and feature dimensions of each layer are listed in \fixme{Appendix 3 (Table 2)}.

\subsection{Semi-Supervised Training with Confidence Learning}
\label{subsec:training-with-cl}
We employ a two-stage training approach integrated with confidence learning (CL) (see the \lf{purple arrows} in Figure~\ref{fig:train_process}).
\textbf{The first stage} involves generating rough predictions for label confidence assessment and noise identification.
\textbf{In the second stage}, the training dataset is refined by filtering out the noise identified through CL, thereby enhancing TNANet's prediction accuracy.

For both of these stages, the TNANet is independently initialized and trained with the same hyperparameters and validation protocols, with the only difference lying in the database used. 

The pseudocode of the entire two-stage training process is summarized in Algorithm~\ref{alg:Framwork}, where steps~\ref{code: init}-\ref{code:trainDone} constitute the training process and step~\ref{code:prediction} predicts sample $i$ at each stage. Step~\ref{code:CLfilter} involves CL, where ${\vec{y^*}_i}^1$ represents the inferred result in the first stage, and ${\vec{y^*}_i}^2$ represents the inferred results in the second stage for sample $i$.

\begin{algorithm}
\caption{Training and Predicting Process of TNANet.} 
\label{alg:Framwork} 
\begin{algorithmic}[1]
\REQUIRE ~~\\ 
The set of true positive samples, \textit{TP};\\
The set of true negative samples, \textit{TN};\\
The set of original uncertain negative samples, $\textit{UN}_1$;\\
TNANet with parameter $\theta$;
\ENSURE ~~\\
Likelihood of being with suicidal ideation for each sample, $\vec{y^*}_i$.

\WHILE {not all ${\vec{y^*}_i}^1$ are calculated}
\label{code:cross1Start}
\STATE Initialize $\theta$ randomly;
\label{code: init}
\STATE Select samples from \textit{TP},\textit{TN}, and $\textit{UN}_1$ to build current dataset as Table \ref{tab:data};
\STATE Unsupervised training of the DBN module;
\STATE Supervised training of the entire network until converged;
\label{code:trainDone}
\STATE Credit ${\vec{y^*}_i}^1$ of samples in the current test set to the total accumulated average.
\label{code:prediction}
\ENDWHILE 
\label{code:crossValidation1}\\

\STATE Calculate the joint distribution matrix $Q_{y^{*}, \tilde{y}}$ and filer noisy samples from $\textit{UN}_1$ according to PBNR method to get $\textit{UN}_2$ with higher purity.\\
\label{code:CLfilter}

\STATE Repeat step \ref{code:cross1Start}-\ref{code:crossValidation1} and replace $\textit{UN}_1$ and ${\vec{y^*}_i}^1$ by $\textit{UN}_2$ and ${\vec{y^*}_i}^2$.
\label{code:crossValidation2}

\RETURN ${\vec{y^*}_i}^1$ and ${\vec{y^*}_i}^2$;
\end{algorithmic}
\end{algorithm}

\subsubsection{Training Process}
\label{subsec:ss-training}
TNANet’s training integrates a self-supervised phase followed by a supervised phase in both of its stages.

\textbf{Self-Supervised Phase:} In this phase, data undergoes bidirectional processing in each RBM unit of the DBN module across three epochs. Parameters are refined based on the reconstruction loss $\mathcal{L}_{con}$, crucial for effective feature encoding. The reconstruction loss is defined as follows:
\begin{align}
    \vec{h_{ij}}^k & = \mathcal{W}_{ij}{\vec{v_{ij}}^k} + b_{ij} \\
    \vec{v_{ij}}^{k+1} & = {{\mathcal{W}_{ij}}^{-1}}{\vec{h_{ij}}^{k}} + b_{ij}^{*} \\
    \mathcal{L}_{con}^{ijk} & = \Vert \vec{v_{ij}}^k - \vec{v_{ij}}^{k+1} \Vert_1
    \label{eq: reconStructionLoss}
\end{align}
where $k\in\{0,1,2\}$ is the iteration index, $i = 1, 2,..., D^m$ is the DBN index for the $i^{th}$ feature, $j\in\{1,2\}$ is the RBM index,  $\mathcal{W}_{ij}$ is the weight matrix of \textit{RBM-j} of the $i^{th}$ DBN (${\mathcal{W}_{ij}}^{-1}$ is its inverse), and $b_{ij}$, $b_{ij}^*$ are the forward and backward biases of \textit{RBM-j} of the $i^{th}$ DBN. 

\textbf{Supervised Phase:} In the supervised learning phase, the network switches to a unidirectional training mode. Data sequentially passes through the DBN, convolution, and classification modules. The cross-entropy loss is computed right after the softmax layer, ensuring effective learning and label prediction accuracy. The loss is defined as~\ref{eq: nickiCrossEntropy}:
\begin{equation}
     \mathcal{L} = - \vec{y^{*}_i}[{\tilde{y}_i}] + log\sum\limits_{j=0}^{1}{e^{\vec{y^{*}_i}[j]}}
\label{eq: nickiCrossEntropy}
\end{equation}
This loss calculation facilitates the backpropagation of errors and fine-tuning of the network parameters.

\subsubsection{Noise Filtering by Confidence Learning}
\label{subsec:cl}
TNANet incorporates CL as a crucial step in its training process, encompassing probability prediction, joint distribution computation, and noise filtering. Each training stage begins with a selection of samples, including \textit{TP}, \textit{TN},  and approximately 2,000 \textit{UN} samples (as defined in Section~\ref{sec:labels}). After cross-validation, model-inferred probabilities are used to gauge \textit{label confidence} ($\vec{y^{*}_i}[{\tilde{y}_i}]$) for each sample, aiding in identifying potential noise.

The process employs the PBNR method~\cite{Northcutt2021CL} to quantify the amount of noise (${ \mathcal{N} }_{noise}$) in the \textit{UN} sample set. Samples with the lowest \textit{label confidence} scores are then excluded, refining the training set for the subsequent stage. This CL-based approach is anticipated to enhance TNANet's performance, especially in the second training stage with a more purified dataset.

Detailed implementations of the CL method are described in \fixme{Appendix 2}.

\section{Experiments}
\label{sec:results}
This section analyzes the classification results of TNANet and its comparison with existing methods (Section~\ref{subsec:comparison}), followed by ablation studies to evaluate key components (Section~\ref{subsec:ablation}), and physiological insights (Section~\ref{subsec:implications}).

\subsection{Preliminary}
\label{subsec:sys-setting}
\noindent\textbf{Datasets.}
Our experiments utilize our proposed dataset (as introduced in Section~\ref{sec:data}) along with three binary-classification EEG datasets from the UEA repository~\cite{bagnall2018uea}: EyesOpenShut, FingerMovements, and SelfRegulationSCP1. These datasets were selected based on their task relevance and similarity in sample size with our proposed dataset.

\noindent\textbf{Dataset Splitting.}
% Our proposed Dataset.
For our dataset, each validation fold involves the random selection of 12 samples for the \textbf{testing set}, which includes 6 from the \textit{TP} set and 6 from the \textit{TN} set (see Table~\ref{tab:data}).
\fixme{Samples with label uncertainty (\textit{UN}) are excluded from the testing set to ensure accurate evaluation metrics.} 
For further validation in Section~\ref{subsec:ablation}, \fixme{3} \textit{TP} samples are randomly selected out of the validation process from the \textit{TP} set. Consequently, the remaining \fixme{21} \textit{TP} and \fixme{15} \textit{TN} samples form the \textbf{training set}. To incorporate data diversity and address label imbalance, \fixme{6} \textit{UN} samples are also included in the training set, assigned negative labels similar to \textit{TN} samples. \fixme{This inclusion strategy is based on the understanding that while most \textit{UN} samples likely do not exhibit suicidal ideation, they are also informative for representation learning. Furthermore, for those \textit{UN} samples potentially containing suicidal ideation, our method can recognize and leverage their influence through weighted filtering.}

% Public Datasets.
For the three public datasets, the original dataset is partitioned into clean and noisy segments (see Table~\ref{tab:dataPublic}). The definition of the noisy segment is consistent with our \textit{UN}, and it is obtained by introducing symmetric label noise~\footnote{Different from our proposed PPG datasets, for the public datasets, this segment maintains its original labels (not all equals to 0), which are then randomly shuffled to introduce noise.}. During validation, the training data in each fold contains noise with the same ratio as the noisy segment. The noisy segment is added to the training data but excluded from the testing set. This approach emulates realistic conditions for noisy learning.

% ===
\begin{table}[ht]
\caption{The numbers of samples in each fold of our PPG dataset.}
\centering
\resizebox{\linewidth}{!}
{
\begin{tabular}{c|c|c|c}
\hline
& \# of \textit{TP}  & \# of \textit{TN} & \# of \textit{UN}
\\ \hline
Training dataset  & \fixme{$21$} & \fixme{$15$} & \fixme{6}
\\ \hline
Test dataset & 6 & 6 & 0
\\ \hline
\end{tabular}
}
\label{tab:data}
\end{table}
% -
\begin{table}[ht]
\caption{The numbers of samples in each fold in the three public datasets. $|\mathcal{D}|$ denotes the size of the dataset.}
\renewcommand\arraystretch{1.5}
\centering
\resizebox{\linewidth}{!}
{
\begin{tabular}{c|c|c}
\hline
& \# of \textit{clean segment} & \# of \textit{noisy segment}
\\ \hline
Training dataset  & \fixme{$\frac{4}{9} |\mathcal{D}|$} & \fixme{$\frac{4}{9} |\mathcal{D}|$}
\\ \hline
Test dataset & \fixme{$\frac{1}{9} |\mathcal{D}|$} & 0
\\ \hline
\end{tabular}
}
\label{tab:dataPublic}
\end{table}

% ===

\noindent\textbf{Baselines.} 
We compare our method with state-of-the-art deep learning architectures and recent advanced noisy data training algorithms.
\textbf{Deep Learning (DL) Methods.} We incorporate DL methods, including \textit{EEGNet}~\cite{lawhern2016eegnet}, \textit{ShallowConvNet}~\cite{schirrmeister2017deep}, and \textit{DGCNN}~\cite{song2018eeg-dgcnn}, \jwq{renowned} for their performance in EEG-based tasks.
\textbf{Noisy Learning Algorithms.}
To specifically address noise handling, we compare our model with recent noisy learning studies, including \textit{Co-teaching}~\cite{han2018co},  \textit{DivideMix}~\cite{li2020dividemix}, \textit{SREA}~\cite{castellani2021estimating}, and \textit{CTW}~\cite{ma2023ctw}. 

\noindent\textbf{Implementation Details.} 
For TNANet and all comparative deep learning (DL) models, we utilize the Adam optimizer with an initial learning rate of 0.001. The maximum training epoch is set at 100, fitting the constraints of our dataset and model complexity. Model parameters are constrained by cross-entropy loss $\mathcal{L}$, defined in Equation \ref{eq: nickiCrossEntropy}. The output layer of each method consists of two units, representing the probability vector $\vec{y^{*}}$.
For noisy learning algorithms, we adhere to the default settings provided in their respective original implementations.
The performance is evaluated using classification \textit{accuracy} and \textit{F1-score}. 
The subject-independent five-fold cross-validation evaluation strategy is employed.
For TNANet, data splitting follows Tables~\ref{tab:data} and~\ref{tab:dataPublic}, in accordance with the five-fold validation approach.
In evaluations on the three public EEG datasets, symmetrical artificial noise is introduced at a ratio of \fixme{0.3}. For TNANet, we adjust the hyperparameters $D^m$ and $T$ to match the dimensionality and time length of each dataset.
Further details on the implementation of baseline deep-learning models and noisy learning strategies are shown in \fixme{Appendix 4}.

\subsection{Comparison with State-of-the-Arts}
\label{subsec:comparison}
\noindent\textbf{Overall Analysis.} The TNANet shows a notable enhancement in performance across the four datasets when compared with the baseline models. Specifically, TNANet exhibits an average accuracy improvement of \fixme{9.00}\% and an average F1 score improvement of \fixme{9.34}\%.
When comparing TNANet with the second best-performing model, DGCNN, TNANet still maintains an advantage with an accuracy improvement of \fixme{3.55}\% and an F1 score improvement of \fixme{2.52}\%, demonstrating its robustness and adaptability in various EEG-based tasks and its potential for broader applications within the field of mental health monitoring.

\noindent\textbf{Our PPG Dataset.}
As shown in Table~\ref{tab:baselines}, TNANet surpasses the average \jwq{accuracy across the baseline models} by \fixme{5.62}\%, achieving an accuracy of \fixme{63.33}\%. Furthermore, it attains an F1 score of \fixme{61.55}\%, exceeding baseline models by \fixme{2.58}\%. 

\noindent\textbf{Public EEG Datasets.} TNANet's performance on the three public EEG datasets demonstrates its generalizability:
\begin{itemize}
    \item On \textit{EyesOpenShut}, TNANet's accuracy improves by \fixme{5.00}\% and the F1 score increases by \fixme{2.43}\% compared to the second-best model.
    \item For \textit{FingerMovements}, TNANet demonstrates a marginal accuracy increase of \fixme{0.55}\% compared to the second-best EEGNet, along with a negligible F1 score difference of \fixme{0.02}\%.
   \item On \textit{SelfRegulationSCP1}, TNANet achieves a higher accuracy by \fixme{2.41\%} and an F1 score enhancement of \fixme{1.09}\% compared to the second-best DGCNN.
\end{itemize}

TNANet's \jwq{enhanced} accuracy and F1 scores reinforce its value in mental health monitoring endeavors.

Moreover, an evaluation of the average model performance per dataset (see the `Mean' row in Table~\ref{tab:baselines}) indicates that our proprietary dataset achieves higher accuracies compared to two of the public datasets. This further validates the effectiveness of our PPG dataset, establishing it as a reliable source for suicidal ideation analysis.

\begin{table*}[ht!]
\caption{Accuracies and F1-scores of Comparative Methods.
`Mean-Baselines' represents the average performance of all models excluding TNANet, while `Mean' reflects the overall average including TNANet.
The `Average' column presents each model's average performance across the four datasets.}
\centering
\renewcommand\arraystretch{1.4}
\resizebox{\linewidth}{!}
{
\begin{tabular}{c|cc|cc|cc|cc|cc}
\hline
\multirow{2}{*}{\textbf{Model\_Name}} & \multicolumn{2}{c|}{\textbf{EyesOpenShut}}                   & \multicolumn{2}{c|}{\textbf{FingerMovements}}                & \multicolumn{2}{c|}{\textbf{SelfRegulationSCP1}}             & \multicolumn{2}{c|}{\textbf{Ours}}                           & \multicolumn{2}{c}{\textbf{Average}}                          \\ \cline{2-11}
                                      & Accuracy                     & F1 Score                     & Accuracy                     & F1 Score                     & Accuracy                     & F1 Score                     & Accuracy                     & F1 Score                     & Accuracy                     & F1 Score                           \\ \hlineB{5}
\textbf{CTW}                          & 0.5143 $\pm$ 0.0821          & 0.5054 $\pm$ 0.0789          & 0.5133 $\pm$ 0.0766          & 0.5116 $\pm$ 0.0770          & 0.7355 $\pm$ 0.0426          & 0.7334 $\pm$ 0.0419          & 0.5424 $\pm$ 0.0833          & 0.5295 $\pm$ 0.0833          & 0.5764  $\pm$ 0.0712          & 0.5700  $\pm$ 0.0703          \\
\textbf{SREA}                         & 0.4857 $\pm$ 0.1071          & 0.4794 $\pm$ 0.1095          & 0.5165 $\pm$ 0.0204          & 0.4382 $\pm$ 0.0788          & 0.6918 $\pm$ 0.1097          & 0.6561 $\pm$ 0.175           & 0.5455 $\pm$ 0.1288          & 0.5431 $\pm$ 0.1282          & 0.5599  $\pm$ 0.0915          & 0.5292  $\pm$ 0.1229          \\
\textbf{DivideMix}                    & 0.4824 $\pm$ 0.1952          & 0.4735 $\pm$ 0.1895          & 0.4653 $\pm$ 0.0566          & 0.4459 $\pm$ 0.0474          & 0.6830 $\pm$ 0.0466          & 0.6752 $\pm$ 0.0473          & 0.5742 $\pm$ 0.1009          & 0.5511 $\pm$ 0.1046          & 0.5512  $\pm$ 0.0998          & 0.5364  $\pm$ 0.0972          \\
\textbf{DGCNN}                        & 0.6000 $\pm$ 0.0897          & 0.5282 $\pm$ 0.2648          & 0.5639 $\pm$ 0.0411          & 0.6087 $\pm$ 0.0408          & 0.8148 $\pm$ 0.0512          & 0.8208 $\pm$ 0.0470          & 0.5750 $\pm$ 0.0718          & 0.6530 $\pm$ 0.0841          & 0.6384  $\pm$ 0.0635          & 0.6527  $\pm$ 0.1092          \\
\textbf{ShallowConvNet}               & 0.5786 $\pm$ 0.0704          & 0.5969 $\pm$ 0.1717          & 0.5333 $\pm$ 0.0313          & 0.5428 $\pm$ 0.1782          & 0.6093 $\pm$ 0.0680          & 0.6313 $\pm$ 0.0816          & 0.5917 $\pm$ 0.0889          & \textbf{0.6932 $\pm$ 0.0240} & 0.5782  $\pm$ 0.0647          & 0.6161  $\pm$ 0.1139          \\
\textbf{Co-teaching}                  & 0.5154 $\pm$ 0.1020          & 0.5031 $\pm$ 0.1079          & 0.4580 $\pm$ 0.0625          & 0.4486 $\pm$ 0.0598          & 0.6693 $\pm$ 0.0302          & 0.6599 $\pm$ 0.0287          & 0.5939 $\pm$ 0.0937          & 0.5755 $\pm$ 0.0922          & 0.5592  $\pm$ 0.0721          & 0.5468  $\pm$ 0.0722          \\
\textbf{EEGNet}                       & 0.6000 $\pm$ 0.1662          & 0.5874 $\pm$ 0.1934          & 0.5681 $\pm$ 0.0676          & \textbf{0.6431 $\pm$ 0.0399} & 0.7130 $\pm$ 0.1648          & 0.7449 $\pm$ 0.1430          & 0.6167 \jwq{$\pm$} 0.0929           & 0.5823 \jwq{$\pm$} 0.1739  & 0.6245  $\pm$ 0.1229          & 0.6394  $\pm$ 0.1376          \\
\textbf{Mean-Baselines}                & 0.5395 $\pm$ 0.1161          & 0.5248 $\pm$ 0.1594          & 0.5169 $\pm$ 0.0509          & 0.5198 $\pm$ 0.0746          & 0.7024 $\pm$ 0.0733          & 0.7031 $\pm$ 0.0806          & 0.5771 $\pm$ 0.0943          & 0.5897 $\pm$ 0.0986          & 0.5840 $\pm$ 0.0837           & 0.5844 $\pm$ 0.1033           \\
\textbf{TNANet (ours)}                & \textbf{0.6500 $\pm$ 0.0772} & \textbf{0.6212 $\pm$ 0.1680} & \textbf{0.5736 $\pm$ 0.0259} & 0.6429 $\pm$ 0.0251          & \textbf{0.8389 $\pm$ 0.0510} & \textbf{0.8317 $\pm$ 0.0593} & \textbf{0.6333 $\pm$ 0.0865} & 0.6155 $\pm$ 0.1695          & \textbf{0.6740  $\pm$ 0.0602} & \textbf{0.6778  $\pm$ 0.1055} \\ \hline
\textbf{Mean}                         & 0.5533  $\pm$ 0.1112         & 0.5369    $\pm$ 0.1605       & 0.5240  $\pm$ 0.0478         & 0.5352    $\pm$ 0.0684       & 0.7195  $\pm$ 0.0705         & 0.7192    $\pm$ 0.0780       & 0.5841  $\pm$ 0.0934         & 0.5929    $\pm$ 0.1075       & 0.5952  $\pm$ 0.0807          & 0.5960    $\pm$ 0.1036        \\ 
 \hline
\end{tabular}
}
\label{tab:baselines}
\end{table*}

\subsection{Ablation Study}
\label{subsec:ablation}
To evaluate the design effectiveness of TNANet, especially the self-supervised DBN training and noise filtration via confidence learning detailed in Section~\ref{subsec:cl}, we conducted \jwq{ablation studies} on our PPG dataset.

\textbf{\fixme{Self-supervised Training} of DBN.}
The self-supervised training phase of the DBN module, targeted at minimizing reconstruction loss, effectively reduces feature redundancy and preserves critical features, essential for handling noisy data and reducing overfitting risks. We compare three training variants: full self-supervised training, self-supervision with only \textit{UN} samples, and no self-supervised phase. The results are depicted in Table~\ref{tab:DBNpreTrain}, showing a distinct hierarchy in performance: full self-supervised training yields the best results in accuracy and F1 score, followed by self-supervision with only \textit{UN} samples, and the lowest performance is observed when the self-supervised phase is entirely excluded. This gradation in outcomes distinctly highlights the critical role of self-supervised training in the DBN module, significantly contributing to the robust and effective performance of TNANet.

\begin{table}[H]
\renewcommand\arraystretch{1.3}
\caption{Accuracies and F1-scores of different conditions for the self-supervised training phase.}
\centering
\resizebox{\linewidth}{!}
{
\begin{tabular}{c|c|c}
   \hline
        \textbf{Condition} & \textbf{Accuracy} & \textbf{F1 Score} \\ \hline
        Without the Phase & 0.5750 $\pm $0.1294 & 0.2784 $\pm$ 0.3691 \\ \hline
        With \textit{UN} Samples & 0.6167 $\pm $0.0989 & 0.5441 $\pm$ 0.2446 \\ \hline
        With Entire Training Set & \textbf{0.6333 $\pm$ 0.0865}  & \textbf{0.6155 $\pm$ 0.1695} \\ \hline
\end{tabular}
}
\label{tab:DBNpreTrain}
\end{table}

\textbf{Effectiveness of CL Method.}
The two-stage training process of TNANet, with an intermediate noise filtration via CL, shows a significant improvement in classification results. The comparative performance before and after applying CL demonstrates the method's effectiveness in identifying and discarding noisy samples, as indicated in Table~\ref{tab:ablation-CL}. 

\begin{table}[H] \scriptsize
\renewcommand\arraystretch{1.1}
\caption{Accuracies and F1-scores in cross-validation before and after CL method.}
\centering
\resizebox{\linewidth}{!}
{
 \begin{tabular}{c|c|c}
    \hline
        \textbf{Stage} & \textbf{Accuracy} & \textbf{F1 Score} \\ \hline
        Before CL & 0.5833 $\pm$ 0.0929 & 0.5507 $\pm$ 0.2690  \\ \hline
        After CL & \textbf{0.6333 $\pm$ 0.0865}  & \textbf{0.6155 $\pm$ 0.1695} \\ \hline
\end{tabular}
}
\label{tab:ablation-CL}
\end{table}

\textbf{Validation of Detected Noise.}
Following the data partitioning strategy described in Section~\ref{subsec:sys-setting}, a subset of \textit{TP} samples is intentionally excluded from the training and testing phases. \jwq{In} each cross-validation fold, these \textit{TP} samples undergo the same predictive analysis as the \textit{UN} samples not included in the training. After cross-validation, these samples receive a probabilistic suicidal ideation ranking alongside the \textit{UN} samples. The methodology posits that if the \textit{TP} samples systematically rank within the top 200 \footnote{A figure that resonates with the actual suicide rate in prisons and is considered manageable for subsequent in-depth assessment.}, the efficacy of the model's noise detection is affirmed. This procedure is systematically replicated, with separate sets of \textit{TP} samples, to reinforce the model's pragmatic value for pinpointing high-risk prisoners. Detailed in Table~\ref{tab:noise-validation}, these rankings substantiate the model's operational effectiveness, fulfilling the practical objectives depicted in Section~\ref{sec:introduction}.

\begin{table}[ht] \scriptsize
\caption{Ranks of the 3-Mixed \textit{TP} Samples.}
\centering
\begin{tabular}{ccccc}
\toprule
\textbf{No.   Validation} & Prisoner ID & Rank & Average                & Overall Average        \\
\midrule
\multirow{3}{*}{1}        & A           & 0    & \multirow{3}{*}{39.33} & \multirow{9}{*}{\normalsize{\textbf{41.33}}} \\
                          & B           & 17   &                        &                        \\
                          & C           & 101  &                        &                        \\ \\
\multirow{3}{*}{2}        & D           & 9    & \multirow{3}{*}{58.33} &                        \\
                          & E           & 59   &                        &                        \\
                          & F           & 107  &                        &                        \\ \\
\multirow{3}{*}{3}        & G           & 0    & \multirow{3}{*}{26.33} &                        \\
                          & H           & 28   &                        &                        \\
                          & I           & 51   &                        &    
                          
                  \\
\bottomrule     
\end{tabular}
\label{tab:noise-validation}
\end{table}

\subsection{Physiological Insights and Implications}
\label{subsec:implications}
TNANet unravels critical physiological features for suicidal ideation prediction (\fixme{Appendix 5, Table 3}), aligning with empirical observations. Key indicators like pNN20, S, SDNN, and SD1 correlate with lower HRV, consistent with prior studies~\cite{chang2017relationships,adolph2018high,tsypes_resting_2018}. The model identifies novel features, including PPG signal amplitude, showcasing the potential of neural networks to capture subtle physiological fluctuations.

\section{Conclusion}
\label{sec:conclusion}
This study introduces the Temporal-Noise-Aware
%-Net 
\dxx{Neural Network} (TNANet), an innovative approach that leverages confidence learning to address the challenge of noisy labels in the physiological signals-based suicidal ideation prediction. A key contribution of our study is the construction of a unique PPG dataset, curated from real-world prison environments. TNANet's effectiveness is further underscored by comparative studies against various deep learning methods and recent noisy learning strategies, across both our proposed dataset and three public datasets. TNANet consistently outperforms these methods, showcasing its potential in mental health monitoring. In the future, we envision extending our approach to multimodal frameworks, integrating a broader spectrum of risk factors to refine predictive accuracy in the realm of mental health.

% \bibliographystyle{named}
% \bibliography{ijcai24}

\clearpage

\section{Supplementary Materials}

\subsection{Extracted Features}
Totally, \fixme{38} features are extracted from the PPG signals, as listed in Table \ref{tab:Features}.

%%%%%%%%%%%%%%%%%%%%%%%%%%%%%%%%%%%%%%%%%%%
% Features Introduction
%%%%%%%%%%%%%%%%%%%%%%%%%%%%%%%%%%%%%%%%%%%%
\hspace{-25pt}
\begin{table*}[!ht] \footnotesize
\centering
    \caption{\fixme{38} features extracted from PPG signals.}
    \begin{tabular}{c|l}
    \hline
    \textbf{Feature Name}  & \textbf{Explanation}
    \\ \hline
    mean T1 & T1 is the time interval between foot(i) and peak(i) in the i-th beat in Figure~\ref{fig:PPG}\\
    mean T2 & T2 is the time interval between peak(i) in the i-th beat and foot(i+1) in the next beat\\
    mean T & T is the time interval between two successive feet in the i-th beat\\
    mean RTR & RTR is the reflection time ratio of rising time to descending time\\
    mean A1 & A1 is the area under the rising waveform from foot(i) to peak(i) in the i-th beat\\
    mean A2 & A2 is the area under the descending waveform from peak(i) in the i-th beat to foot(i+1) in the next beat\\
    mean A & A is the area under the waveform in the whole i-th beat\\
    mean RAR & RAR is the reflection area ratio of rising time to descending time\\
    mean H1 & H1 is the amplitude between foot(i) and peak(i) in the i-th beat\\
    mean H2 & H2 is the amplitude between peak(i) in the i-th beat and foot(i+1) in the next beat\\
    mean RPR & RPR is the ratio of rising time to descending time\\
    mean amplitude & amplitude of filtered PPG\\
    mean IBI & IBI is the time interval between peak(i) in the i-th beat to peak(i+1) in the next beat\\
    stds & standard deviation of \{T1, T2, T, A1, A2, A, H1, H2, RTR, RAR, RPR, amplitude\}\\
    % --- 25 above ---
    SDNN & standard deviation of successive inter-beat interval\\
    RMSSD & root mean square difference of successive inter-beat interval\\
    pNN20 & the proportion of N20 (number of pairs of successive IBI that differ by more than 20ms) divided by the total number of IBI\\	
    pNN50 & the proportion of N50 (number of pairs of successive IBI that differ by more than 50ms) divided by the total number of IBI\\	
    energy & the squared original sequence after filtering\\
    time duration & the duration of the entire stimulation state\\
    bandwidth & the sum of the squared first-order-difference divided by the feature \textit{energy}\\
    time-bandwidth product & {the product of feature \textit{time duration} and \textit{bandwidth}}\\
    % --- 33 above --- newly extracted below ---
    heart rate & {the average number of detected peaks per 60s}\\ 
    entropy & {Shannon Entropy calculated on the detected peaks as in~\cite{dash2009automatic}}\\
    S, SD1, SD2 & {Poincare analysis}\\
    % --- 38 --- 
    \hline
    \end{tabular}
    \label{tab:Features}
\end{table*}
\begin{figure*}[!ht]
    \centering
    \includegraphics[width=0.7\linewidth]{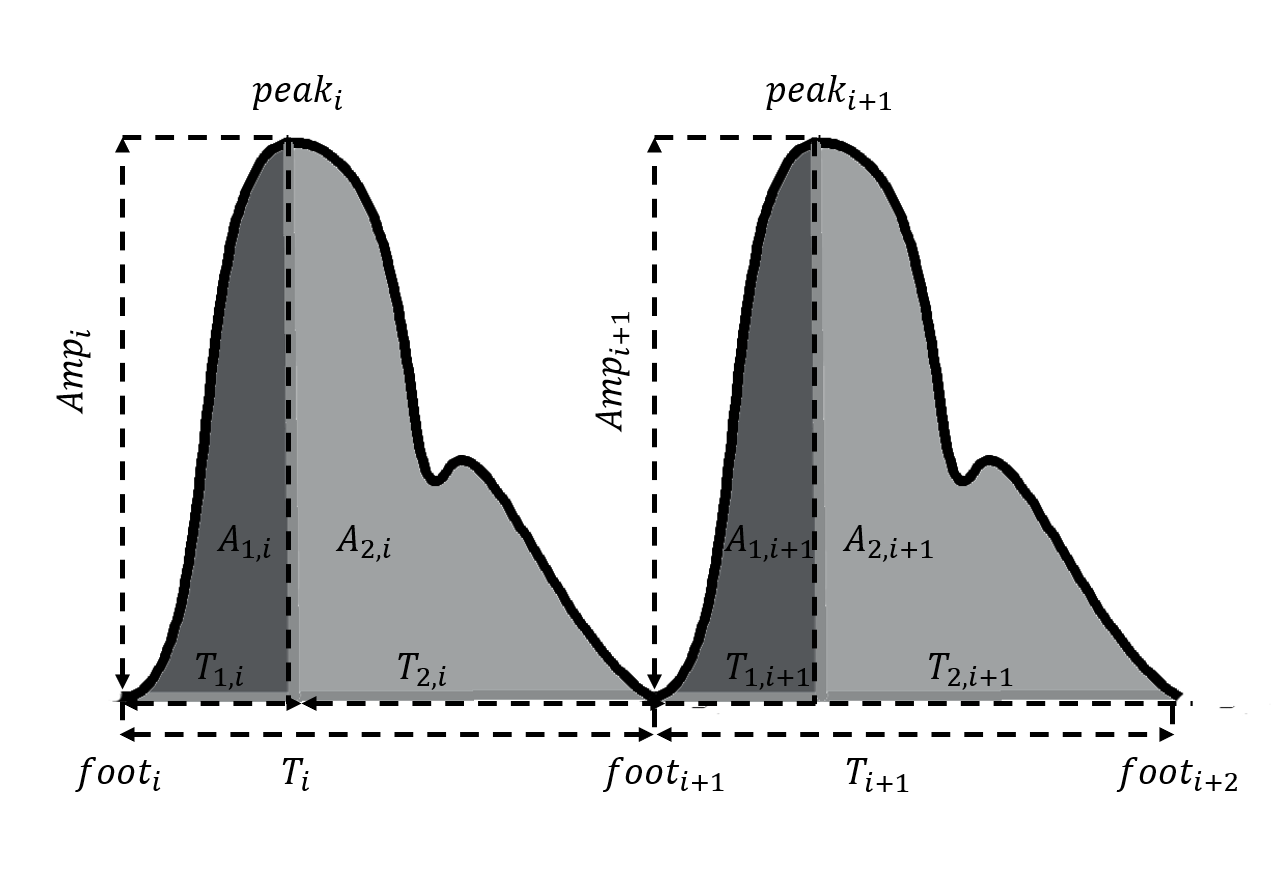}
    \caption{Heartbeats detection in photoplethysmogram (PPG) signals.}
    \label{fig:PPG}
\end{figure*}

\subsection{Preliminary of confidence learning}
The confidence learning (CL) method, proposed by Northcutt et al., is based on statistical rules and is independent of the classifier structure and the training process. The goal of CL is to identify mislabeled samples and remove them from the dataset before training. 
CL focuses on the category-related features of the noise and operates only once to calculate the $2\times 2$ joint distribution matrix $\hat{Q}_{\widetilde{y}, y^{+}}$ between the given ($\widetilde{y}$) and estimated ($y^{+}$) labels. Specifically, the element $\hat{Q}_{\widetilde{y}=0, y^{+}=1}$ represents the ratio of samples with negative given labels but assumed positive by machine learning methods, which indicates the ratio of hard negative samples.
 
The CL method mainly contains three phases, i.e. (1) joint distribution estimation, (2) noise filtering, and (3) retraining on purified samples. These three phases will be introduced sequentially following the notations.

\textbf{Notations}
For labels, $\tilde{y}_k \in \{0, 1\}$ represents the given label of sample $k$ after the labeling process as introduced in Section~\ref{sec:labels} in the main paper. $\vec{y^{*}_k} \in \mathbb{R}^2$ is a 2-dim vector, where $\vec{y^{*}_k}[0]$ stands for the output probability of being negative and $\vec{y^{*}_k}[1]$ stands for the output probability of being positive.

\textbf{Joint Distribution Estimation.}
In this phase, only \textit{TN} and \textit{TP} samples are used. Firstly, the ``label-confidence'' $\widehat{p}_k$ is calculated by $\widehat{p}_k = \vec{y^{*}_k}[{\tilde{y}_k}]$, which is the probability of sample $k$ to be correctly labeled. Heuristically, a sample with low label confidence is likely to have been mislabeled. 

Let $X_0$ and $X_1$ respectively represent the \fixme{number} of \textit{TN} and \textit{TP} samples. Then we calculate the mean label-confidence \fixme{$t_c$} of each category $c\in \{0,1\}$:\\
    \begin{equation}
        t_c=\frac{1}{\fixme{X_c}}
        {
            \sum\limits_{{\widetilde{y}_k}=c}
            {\widehat{p}_k}
        }
    \label{eq: categoryConfidence}
    \end{equation}
%Equation~\ref{eq:noiseDEFINE} defines the \textit{CL label $y^{+}$} of a sample. 
Then the estimated label $y^{+}$ for CL analysis is calculated by Equation~\ref{eq:CLlabel}.
Specifically, $y^{+}_k = c$ means that the output probability of sample k under category c is above c's confidence (i.e., $t_c$ in Equation~\ref{eq: categoryConfidence}),
and $\vec{y^{*}_k}[c] \ge \vec{y^{*}_k}[1-c]$ if the conditions $\vec{y^{*}_k}[0] \ge t_0$ and $\vec{y^{*}_k}[1] \ge t_1$  are simultaneously satisfied. Otherwise, if neither of the two conditions is satisfied, $y_k^{+}$ is set to `null', and sample $k$ is neglected. Then similar to the calculation of the confusion matrix, the confidence joint matrix $C_{\widetilde{y}, y^{+}}$ is computed using Equation~\ref{eq:noiseDEFINE}.
Such a definition considers the class characteristics, preventing class imbalance due to low categorical confidence. At the same time, for samples with low output probability across categories, the method does not count them, avoiding the interference of categorically inconspicuous samples.
\small{
    \begin{equation}
        y_k^{+} = {\mathop{\arg\max}\limits_{c\in\{0,1\}:{\hat{p}}_k \ge t_c}}{\vec{y^{*}}_k[c]}
        \label{eq:CLlabel}
    \end{equation}
}

\small{
    \begin{equation}
        {\fixme{C}}_{\widetilde{y}=i, y^{+}=j} := {\fixme{C}}_{\widetilde{y}, y^{+}}[i][j] =
        |\left \{
        {sample}_k |\quad
        \widetilde{y}_k=i, y_k^{+}=j
        \right \}|
        \label{eq:noiseDEFINE}
    \end{equation}
}

Based on $C_{\widetilde{y}, y^{+}}$, we calculate the joint distribution ${\hat{Q}}_{\widetilde{y}, y^{+}}$ as:
\iffalse
    $$
    {\hat{Q}}_{\widetilde{y}=i, y^{+}=j} =
    $$
    \begin{equation}
        \frac
        {\frac{C_{\widetilde{y}=i,y^{+}=j}}{\sum_{j\in\{0,1\}{C_{\widetilde{y}=i,y^{+}=j}}}\cdot\fixme{X_i}}}
        {
            \sum\limits_{i,j\in\{0,1\}}
            {(\frac
                {C_{\widetilde{y}=i,y^{+}=j}}
                {\sum_{j\in\{0,1\}{C_{\widetilde{y}=i,y^{+}=j}}}\cdot\fixme{X_i}}
             )}
        }
    \end{equation}
\fi
    
\begin{equation}
    {\hat{Q}}_{\widetilde{y}=i, y^{+}=j}=\frac
        {\frac{C_{\widetilde{y}=i,y^{+}=j}}{\sum_{j\in\{0,1\}{C_{\widetilde{y}=i,y^{+}=j}}}\cdot\fixme{X_i}}}
        {
            \sum\limits_{i,j\in\{0,1\}}
            {(\frac
                {C_{\widetilde{y}=i,y^{+}=j}}
                {\sum_{j\in\{0,1\}{C_{\widetilde{y}=i,y^{+}=j}}}\cdot\fixme{X_i}}
             )}
        }
    \end{equation}   
    
\textbf{Noise Filtering.}
According to the PBNR method proposed in~\cite{Northcutt2021CL}, we select $|\textit{UN}| \times \fixme{\hat{Q}}_{\widetilde{y}=0,y^{+}=1}$ ($|\textit{UN}|$ is the number of \textit{UN} samples) \textit{UN} samples out with the lowest label-confidence (i.e. samples with the max margin $\vec{y^{*}_k}[1]-\vec{y^{*}_k}[0]$).

\textbf{Retraining on Purified Samples.} After filtering samples with high probabilities of being mislabeled, we obtain a purified dataset, on which we perform the second stage cross-validation of our TNANet.

\subsection{Detailed Architecture of TNANet}
Table~\ref{tab:DBNconv-architecture} shows the details of each layer of TNANet, where $D^m$ is the number of extracted features from PPG signals, T is the number of temporal windows, $H_1$ is the number of units in the hidden layer of RBM-1, $H_2$ is the number of units in the hidden layer of RBM-2, F is the number of filters (set to 16), P$=\min({H_2}//4, 8)$ is a pooling size in the second block of the convolution module, and C is the number of classification categories (set to 2), respectively.
%%%%%%%%%%%%%%%%%%%%%%%%%%%%%%%%%%%%%%%%%%%
% Architecture Of DBN_ConvNet
%%%%%%%%%%%%%%%%%%%%%%%%%%%%%%%%%%%%%%%%%%%%
\begin{table*}[!ht]\small
\caption{Architecture for TNANet.}
\centering
\begin{tabular}{cccccc}
\toprule[2pt]
\textbf{Module}                          & \textbf{Layer}  & \textbf{\#instance\fixme{s}} & \textbf{\#filters} & \textbf{size}                 & \textbf{output}              \\
\midrule[1.5pt]
\textbf{}                                & Input           &                     &                    &                               & (1, $D^m$, T)                \\
\textbf{}                                & Split           & $D^m$               &                    &                               & (1,  T)                      \\
\midrule[0.5pt]
\multirow{2}{*}{\textbf{DBN}}            & RBM-1            & $D^m$               &                    &                               & (1, $H_1$)                  \\
                                         & RBM-2            & $D^m$               &                    &                               & (1, $H_2$)                  \\
\midrule[0.5pt]
\textbf{}                                & Concatenate     &                     &                    &                               & (1, $D^m$, $H_2$)           \\
\midrule[0.5pt]
\multirow{10}{*}{\textbf{Convolution}}   & DepthwiseConv2D &                     & F              & ($D^m$, 1)                    & (F, 1, $H_2$)               \\
                                         & BatchNorm       &                     &                    &                               & (F, 1, $H_2$)               \\
                                         & ELU             &                     &                    &                               & (F, 1, $H_2$)               \\
                                         & AveragePool2D   &                     &                    & (1,4)                         & (F, 1, $H_2$//4)            \\
                                         &                 &                     &                    &                               &                              \\
                                         & ZeroPad2D       &                     &                    & ((F - 1) // 2, F // 2, 0, 0)  & (F, 1, $H_2$//4 +F-1 )      \\
                                         & SeparableConv2D &                     & F              & (1, F)                    & (F, 1, $H_2$//4)            \\
                                         & BatchNorm       &                     &                    &                               & (F, 1, $H_2$//4)            \\
                                         & ELU             &                     &                    &                               & (F, 1, $H_2$//4)            \\
                                         & AveragePool2D   &                     &                    & (1, P)            & (F, 1, $H_2$//(4$\cdot$P)) \\
\midrule[0.5pt]
\multirow{3}{*}{\textbf{Classification}} & Flatten         &                     &                    &                               & (F $\cdot$ $H_2$//(4$\cdot$P))   \\
                                         & Linear          &                     &                    & (F $\cdot$ $H_2$//(4 $\cdot$ P), C) & (2)                          \\
                                         & Softmax         &                     &                    &                               & (2)       \\        
\bottomrule[2pt]                                         
\end{tabular}
\label{tab:DBNconv-architecture}
\end{table*}

\subsection{Implementation Details of Comparative Methods.}
% TODO: implementation details of EEGNet
\textbf{Deep Learning (DL) Methods.} The implementation details of each neural network model remain consistent with the original settings in the temporal domain, while the weighting capabilities in the spatial domain are leveraged for automatic feature selection. The specific implementation is elaborated as follows:
\begin{itemize}
    \item \textbf{EEGNet.} EEGNet commences with a zero-padding layer, followed by a temporal convolutional layer (filter size: $1 \times 64$, kernel number: $F1=8$), and a depth-wise convolutional layer (filter size: $D^m \times 1$, kernel number: $F2=16$, groups number: $F1$). This is succeeded by separable convolution (two consecutive layers, the first with filter size: $1 \times 16$, kernel number: $F2$, groups number: $F2$, and the second with filter size: $1 \times 1$, kernel number: $F2$). The network includes batch normalization, ELU activation, average pooling, and dropout. The final output is through a fully connected layer and softmax function, yielding classification probabilities.

    \item \textbf{ShallowConvNet.} This model includes a temporal convolutional layer (kernel size: $1\times 10$, kernel number: $F1=25$), followed by a spatial convolutional layer (kernel size: $1\times D^m$, kernel number: $F2=25$). Subsequently, a pooling layer (temporal size: $1 \times 3$, spatial size: $1\times 3$) and an activation layer are incorporated, simulating logarithmic power computation. This configuration effectively analyses temporal and spatial EEG signal aspects.

    \item \textbf{DGCNN.} In DGCNN, extracted features are considered as graph nodes. The network begins with spatial domain batch normalization, followed by a 5-level Chebyshev polynomial graph convolution on the adjacency matrix $A$, compressing the temporal length from $T$ to 8. Two consecutive fully connected layers are applied thereafter, with dimensions $(D^m \cdot 8) \times 25$ and $25 \times \fixme{C}$. This design enables DGCNN to capture complex relationships among EEG features.
\end{itemize}

\textbf{Noisy Learning Strategies.} These strategies include Co-teaching~\cite{han2018co}, DivideMix~\cite{li2020dividemix}, SREA~\cite{castellani2021estimating}, and CTW~\cite{ma2023ctw}. For each strategy, we adhered to the default hyperparameter settings as specified in their respective original papers and implementations. This ensures a fair and consistent comparison with our TNANet model.

\subsection{Feature Importance Learned by TNANet}
Table~\ref{tab: FeatureImportance} shows the outcomes in the descending order of the total \fixme{38} features, regarding their contribution to the overall prediction performance. The order is determined based on the \fixme{absolute} mean value of the convolutional weights from the first convolutional block of TNANet, as outlined in Table~\ref{tab:DBNconv-architecture}.

%%%%%%%%%%%%%%%%%%%%%%%%%%%%%%%%%%%%%%%%%%%
% Feature Importance 
%%%%%%%%%%%%%%%%%%%%%%%%%%%%%%%%%%%%%%%%%%%%
\begin{table}[!ht]
\caption{Feature importance by TNANet.}
    \centering
    \begin{tabular}{c|c}
    \hlineB{5}
        \textbf{Rank} & \textbf{Feature}\\ \hlineB{3}
        1 & mean H2 \\ \hline
        2 & pNN20 \\ \hline
        3 & std T2 \\ \hline
        4 & std H1 \\ \hline
        5 & S \\ \hline
        6 & mean RTR \\ \hline
        7 & time duration \\ \hline
        8 & std T \\ \hline
        9 & std amplitude \\ \hline
        10 & bandwidth \\ \hline
        11 & std RPR \\ \hline
        12 & SDNN \\ \hline
        13 & SD1 \\ \hline
        14 & mean RPR \\ \hline
        15 & mean T \\ \hline
        16 & mean A2 \\ \hline
        17 & mean amplitude \\ \hline
        18 & mean RAR \\ \hline
        19 & pNN50 \\ \hline
        20 & std A \\ \hline
        21 & std RTR \\ \hline
        22 & std H2 \\ \hline
        23 & std RAR \\ \hline
        24 & SD2 \\ \hline
        25 & time-bandwidth product \\ \hline
        26 & IBI \\ \hline
        27 & mean T2 \\ \hline
        28 & mean A \\ \hline
        29 & mean T1 \\ \hline
        30 & std A1 \\ \hline
        31 & std A2 \\ \hline
        32 & entropy \\ \hline
        33 & mean A1 \\ \hline
        34 & std T1 \\ \hline
        35 & mean H1 \\ \hline
        36 & energy \\ \hline
        37 & RMSSD \\ \hline
        38 & heart rate \\ \hlineB{5}
    \end{tabular}
\label{tab: FeatureImportance}
\end{table}

\indent\setlength{\parindent}{2em}

\end{document}